 \documentclass[final,3p,times,twocolumn]{elsarticle}

\usepackage{amssymb}

\biboptions{sort&compress}

\journal{NIM B385(2016)1}

\begin{document}

\begin{frontmatter}



\title{Excitation function of the alpha particle induced nuclear reactions on enriched $^{116}$Cd, production of the theranostic isotope $^{117m}$Sn}


\author[1]{F. Ditr\'oi\corref{*}}
\author[1]{S. Tak\'acs}
\author[2]{H. Haba}
\author[2]{Y. Komori}
\author[3]{M. Aikawa}
\author[1]{Z. Sz\H ucs}
\author[3]{M. Saito}
\cortext[*]{Corresponding author: ditroi@atomki.hu}

\address[1]{Institute for Nuclear Research, Hungarian Academy of Sciences (ATOMKI),  Debrecen, Hungary}
\address[2]{RIKEN Nishina Center, Tokyo, Japan}
\address[3]{Graduate School of Science, Hokkaido University, Sapporo, Japan}

\begin{abstract}
$^{117m}$Sn is one of the radioisotopes can be beneficially produced through alpha particle irradiation. The targets were prepared by deposition of $^{116}$Cd metal onto high purity 12 $\mu$m thick Cu backing. The average deposited thickness was 21.9 $\mu$m. The beam energy was thoroughly measured by Time of Flight (TOF) methods and proved to be 51.2 MeV. For the experiment the well-established stacked foil technique was used. In addition to the Cd targets, Ti foils were also inserted into the stacks for energy and intensity monitoring. The Cu backings were also used for monitoring and as recoil catcher of the reaction products from the cadmium layer. The activities of the irradiated foils were measured with HPGe detector for gamma-ray spectrometry and cross section values were determined. As a result excitation functions for the formation of $^{117m}$Sn, $^{117m,g}$In, $^{116m}$In, $^{115m}$In and $^{115m,g}$Cd from enriched $^{116}$Cd were deduced and compared with the available literature data and with the results of the nuclear reaction model code calculations EMPIRE 3.2 and TALYS 1.8. Yield curves were also deduced for the measured nuclear reactions and compared with the literature.
\end{abstract}


\end{frontmatter}


\section{Introduction}
\label{1}
The radioisotope $^{117m}$Sn has already been suggested to be useful both for diagnostic and for therapeutic purposes at the same time. It is used as palliative agent in the case of bone cancer metastases \cite{1,2,3,4}, intra-tumoral administration in the cases of micro-metastases, leukemia, lymphomas \cite{5}, but also for diagnostic imaging purposes \cite{6}. Its non-oncological use is also known in the case of synovectomy and marrow ablation \cite{5} and also used in treatment of vulnerable plaques \cite{7, 8}, where it has been proved to be unique. $^{117m}$Sn, as a conversion electron emitter radioisotope with its T$_{1/2}$ = 14 days and ECE = 126.82 keV (65.7 \%), 129.36 keV (11.65 \%) and 151.56 keV (26.5\%) \cite{9} most intense conversion electrons, is more preferable therapeutic and palliative agent in the case of bone metastases than the more frequently used $\beta^-$-emitters \cite{2}. Because of its shorter effective range, it does not cause significant damages to the very sensitive bone marrow. Its E$_{\gamma}$ = 158.56 keV (86.4 \%) most intense $\gamma$-line \cite{9} makes it suitable for diagnostic imaging purposes \cite{6}. The recent term “theranostic” radioisotope is used for denomination of such medical radioisotopes. It has also been reported that $^{117m}$Sn is the best choice for commercial production and labeling a large variety of molecules \cite{7}. 
The most common method for $^{117m}$Sn production was the (n,n’$\gamma$) reaction in nuclear reactors on tin target \cite{10}, which produces low specific activity final product \cite{11}. There are several charged particle routes to produce the $^{117m}$Sn radioisotope. By using high energy protons for antimony irradiation, one can produce NCA (no carrier-added) $^{117m}$Sn \cite{10, 12}, which is also proved to be preferable in most of the nuclear medicine applications, but the optimization for low $^{113m}$Sn contamination goes at the expense of the produced $^{117m}$Sn activity and requires enriched target material. The first reports on cyclotron production of $^{117m}$Sn on In and Cd targets was already published in the early 60s \cite{13} without cross section data. Qaim and D\"ohler reported carrier-free $^{117m}$Sn production data on natural cadmium and indium \cite{14} in an extensive energy range up to 140 MeV. This work deals also with target preparation and radiochemistry of the final product. The different production routes are discussed in detail by Qaim too \cite{15}. The $^{117m}$Sn is produced practically on the two highest mass stable cadmium isotopes 114 and 116. $^{117m}$Sn production is also discussed in \cite{16, 17} without giving cross section data.  The most recent paper dealing, among others with $^{117m}$Sn production \cite{18} gives also detailed review of production routes and applications without numerical cross section data. The most proper production route for medical purposes is the alpha particle induced reaction on both natural cadmium and on enriched $^{116}$Cd targets \cite{19}, because the alpha particles are the most proper bombarding particles to excite high spin isomeric levels \cite{18}, as $^{117m}$Sn (11/2$^-$,  see in Table 1). In this case the isomeric ratio is high and the population of the high-spin states is preferred.  Certainly the enriched target gives the purest final product (NCA) with the highest production yield. The goal of this work was to extend the energy range of the existing nuclear data up to 51 MeV and clarify the discrepancies between the existing nuclear data \cite{19, 20}.

\section{Experimental}
\label{2}
Natural cadmium consists of 8 stable isotopes, and the natural abundance of $^{116}$Cd is only 7.47 \%. Our targets were prepared by electrodeposition of enriched $^{116}$Cd ($^{106}$Cd:0.0006, $^{108}$Cd: 0.0003, $^{110}$Cd: 0.0024, $^{111}$Cd: 0.0029, $^{112}$Cd: 0.0049, $^{113}$Cd: 0.0049, $^{114}$Cd: 0.0147, $^{116}$Cd: 0.9707 against the natural composition of $^{106}$Cd:0.0125, $^{108}$Cd: 0.0089, $^{110}$Cd: 0.1251, $^{111}$Cd: 0.1281, $^{112}$Cd: 0.2413, $^{113}$Cd: 0.1222, $^{114}$Cd: 0.2872, $^{116}$Cd: 0.0747) on 12 $\mu$m thick Goodfellow high purity (99.9 \%) copper backing foils. The electro-deposition was performed in alkaline bath using ammonia as complex builder. Hydrazine hydrate helped by anodic depolarization and Brij-35 agent was the surfactant (all analytical grade). The enriched Cd material was purchased from Isoflex Europe. The thicknesses of the Cd layers were determined as 21.9 $\mu$m average. The stacks containing eight electro-deposited cadmium layers on copper backing and three 10.9 $\mu$m thick titanium foils for further monitoring purposes were irradiated at the beam line of the RIKEN Nishina Center K70-MeV AVF cyclotron with an E$_{\alpha}$ = 51.2 MeV alpha beam. The exact energy of the bombarding alpha particles was determined by Time of Flight (TOF) methods and was found to be E$_{\alpha}$ = 51.2 MeV \cite{21, 22}. The samples were irradiated for 2 hours with an average beam current of I$_{\alpha}$ = 50 nA in a Faraday-cup like target holder. The beam current was integrated and recorded during the irradiation and it was proved to be constant within 2 \%. The beam current and energy were checked by re-measuring the $^{nat}$Ti($\alpha$,x)$^{51}$Cr, the $^{nat}$Cu($\alpha$,x)$^{65}$Zn, the $^{nat}$Cu($\alpha$,x)$^{67}$Ga and the $^{nat}$Cu($\alpha$,x)$^{66}$Ga nuclear reaction cross  sections in the whole energy region and by comparing them with the recommended data from the IAEA monitor reaction library \cite{23}. After the irradiation the samples were removed from the irradiation position as quickly as possible, and separated one from another in order to catch the activity of the short-lived radioisotopes too. Two irradiations were performed, one in January and another one in December 2015, respectively, to re-check the decay contribution from $^{117}$In to $^{117m}$Sn, which might have occurred. This contribution was proved to be low but measurable. The samples were measured by using low background HPGe gamma spectrometry systems based on Ortec detectors, nuclear electronics and data acquisition software. The samples were measured in four series, the first one started just after EOB and the last one two weeks later. The enriched Cd samples were measured four times and results for the different radioisotopes were taken from the most appropriate series of measurements. The cross sections of the different reactions were calculated by using the well-established activation formula taking into account the corresponding peak area, irradiation, cooling and measuring times and the nuclear parameters of the reactions and isotopes \cite{9}. 
The uncertainties of the cross sections were determined by taking square root of the sum in quadrature of all individual contributions \cite{24}: particle flux (5\%),  target thickness (3\%), detector efficiency (5\%), nuclear data (3\%), peak area determination  and counting statistics (1-20\%), the overall uncertainty in the final cross section values was 7-20 
While calculating the cross sections, interferences from different sources might occur, that’s why the data were thoroughly selected and corrections were applied if necessary.

\begin{table*}[t]
\tiny
\caption{Nuclear data for the radioisotopes produced \cite{9, 25}}
\centering
\begin{center}
\begin{tabular}{|p{0.6in}|p{0.3in}|p{0.4in}|p{0.15in}|p{0.15in}|p{0.7in}|p{0.3in}|} \hline 
\textbf{Isotope\newline spin\newline level energy(keV)} & \textbf{Half-life\newline } & \textbf{Decay mode\newline \%\newline } & \textbf{E$_{\gamma}$\newline keV\newline } & \textbf{I$_{\gamma}$\newline \%\newline } & \textbf{Contributing reactions\newline on $^{116}$Cd} & \textbf{Threshold\newline MeV\newline } \\ \hline 
$^{117m}$Sn  & 14 d & IT 100  & 156.0 & 2.1 & $\alpha$,3n & 21.11 \\ \cline{6-7} 
11/2$^{-}$\newline 314.58 &  &  & 158.6 & 86.4 & $^{117}$In decay &  \\ \hline 
 
$^{117m2}$In  & 116.2 min & IT 47.1  & 315.3 & 19.1 & $\alpha$,2np & 21.81 \\ \cline{6-7} 
 1/2$^{-}$\newline 315.302&  & ${\beta}^{-}$ 52.9 & ~ & ~ & $^{117}$Cd decay &  \\ \hline 
 
$^{117g}$In  & 43.2 min & ${\beta}^{-}$ 100 & 158.6 & 87 & $\alpha$,2np & 21.49 \\ \cline{6-7}  
 9/2$^{+}$&  &  & 552.9 & 100 & $^{117}$In decay & ~ \\ \hline 

$^{116m}$In & 54.29 min & ${\beta}^{-}$ 100 & 1097.3 & 58.5 & $\alpha$,3np & 21.49 \\ \cline{6-7} 
 5$^{+}$\newline 127.276 &  &  & 1293.6 & 84.8 & ~ &  \\ \hline 

$^{115m}$In & 4.486 h & IT 95  & 336.2 & 45.8 & $\alpha$,4np & 37.92 \\ \cline{6-7}
 1/2$^{-}$\newline 336.244 &  & ${\beta}^{-}$ 5 & ~ & ~ & $^{115}$Cd decay &  \\ \hline 
 
$^{115g}$Cd  & 53.46 h &${\beta}^{-}$ 100 & 527.9 & 27.5 & $\alpha$,3n2p & 38.27 \\ \cline{6-7} 
1/2$^{+}$ &   &  & ~ & ~ & $\alpha$,n$\alpha$ & 9.0 \\ \hline 

$^{115m}$Cd  & 44.56 d & ${\beta}^{-}$ 100 & 933.8 & 2 & $\alpha$,3n2p & 38.45 \\ \cline{6-7} 
11/2$^{-}$\newline 181.05 &  &  & ~ & ~ & $\alpha$,n$\alpha$ & 9.180 \\ \hline 
 
\end{tabular}

\end{center}

\end{table*}




\section{Theoretical nuclear reaction model calculations}
\label{3}
In order to check the trend of our results and to test the prediction capability of the nuclear reaction model codes, calculations for the measured cross sections were made by using the adjusted values of TALYS 1.8 code \cite{26} tabulated in the TENDL-2015 on-line data library \cite{27, 28}. Calculations with the newest version of the EMPIRE code \cite{29}, the EMPIRE 3.2 (Malta) \cite{30}, which incorporates the newest reference input parameters library RIPL-3 \cite{31}, were also presented for comparison. These calculations were made by using the default input parameters and taking into account all possible reactions, including emission of complex particles at the given bombarding energies. The level density fits were corrected after the first run if it was necessary. These comparisons help the developers of the above codes to further improve the theory behind these calculations.

\section{Results}
\label{4}

\subsection{Excitation functions}
\label{4.1}
In this study excitation functions the formation of $^{117m}$Sn, $^{117m}$In, $^{117g}$In, $^{116m}$In, $^{115m}$In, $^{115m}$Cd and $^{115}$Cd were determined and are presented together with the previous results of different authors and the results of the theoretical model code calculations. 

\subsubsection{$^{116}$Cd($\alpha$,3n)$^{117m}$Sn reaction}
\label{4.1.1}
The main goal of this study was the investigation of the production possibility of the medically important $^{117m}$Sn by using high energy alpha particle beams on enriched target material. The only direct route to produce this radioisotope is the ($\alpha$,3n) reaction on $^{116}$Cd, but it can also be produced from the decay of the $^{117m,g}$In mother isotopes. The T$_{1/2}$ = 14 days of the $^{117m}$Sn radioisotope made a convenient measurement possible. Already at 25 MeV the cross section of $^{116}$Cd($\alpha$,3n) reaction is an order of magnitude higher than that of $^{114}$Cd($\alpha$,n) \cite{19} and this ratio increases with the increasing bombarding energy, as well as the enrichment of $^{116}$Cd is more than 2 orders of magnitude higher than that of the $^{114}$Cd, that is why the contribution from $^{114}$Cd is negligible.  The samples were measured several times and both the E$_{\gamma}$ = 156.0 keV (2.113\%) and the E$_{\gamma}$ = 158.6 keV (86.4\%) gamma-lines were used for activity assessment. The data from a late measurement series after a suitable long cooling time were taken, because we had to wait for the decay of the $^{117m}$In (T$_{1/2}$ = 116.2 min), which has a very close gamma-line (E$_{\gamma}$ =158.6 keV (0.87\%)), and $^{117g}$In mother isotope with T$_{1/2}$ = 43.2 min, respectively. Both isomeric states decay finally to $^{117m}$Sn via different decay routes. In such a way the presented data are cumulative. The result of this work is presented in Fig. 1, together with the results of the nuclear reaction model calculations with EMPIRE 3.2 (Malta) and TALYS 1.8 taken from the TENDL-2015 on-line library. To guide the eyes a spline fit on the experimental data was also presented in the figures, if it was necessary.  Our data are in acceptable agreement with the previous results of Adam-Rebeles et al. \cite{19} and Montgomery et al. \cite{20}; however, our data are slightly lower in the 35 to 40 MeV energy region. Our data above 42 MeV are new. The results of the theoretical nuclear reaction model calculations follow the trend of the experimental data, but both underestimate them around the maximum. There is an energy shift between them, but it is difficult to judge, which is shifted, because under 25 MeV TENDL-2015 gives good estimation of the experimental values, but above 38 MeV the estimation of EMPIRE 3.2 is much better.

\begin{figure}
\includegraphics[scale=0.3]{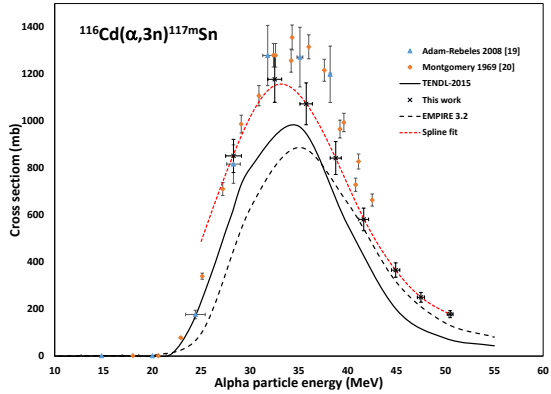}
\caption{Excitation function of the $^{116}$Cd($\alpha$,3n)$^{117m}$Sn nuclear reaction in comparison with the previous literature data and the results of the theoretical nuclear reaction model codes EMPIRE 3.2 and TALYS 1.8 (TENDL-2015)}
\end{figure}

\subsubsection{$^{116}$Cd($\alpha$,x)$^{117m}$In reaction}
\label{4.1.2}
The $^{117m}$In radioisotope can be directly produced through the ($\alpha$,p2n) reaction and indirectly from the decay of its longer-lived $^{117}$Cd mother isotope. The analysis was made by using its independent 315.3 keV (19.1\%) gamma-line by using a measurement series with short cooling time to decrease the contribution from the mother isotope. A correction was also made to separate the contribution of the mother isotope. The previous results of Adam-Rebeles et al. \cite{19} are somewhat lower, the TENDL-2015 follows the trend of the experimental data but an acceptable agreement can only be seen up to 35 MeV. It strongly overestimates above this energy (see Fig. 2). The results of EMPIRE 3.2 strongly increase above 28 MeV and give unacceptable results.

\begin{figure}
\includegraphics[scale=0.3]{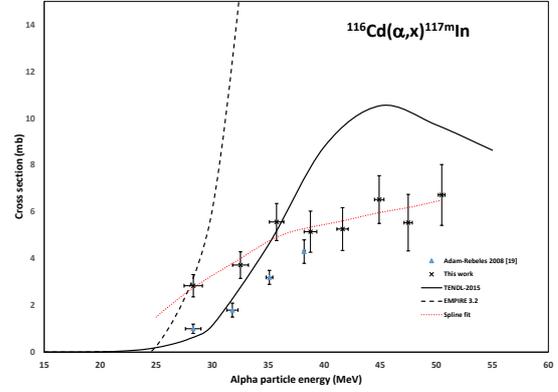}
\caption{Excitation function of the $^{116}$Cd($\alpha$,x)$^{117m}$In nuclear reaction in comparison with the previous literature data and the results of the theoretical nuclear reaction model codes EMPIRE 3.2 and TALYS 1.8 (TENDL-2015)}
\end{figure}

\subsubsection{$^{116}$Cd($\alpha$,x)$^{117g}$In reaction}
\label{4.1.3}
This radioisotope is produced directly in the same nuclear reaction as its isomer, and is also contributed by the decay of the $^{117}$Cd mother isotope. The activities were measured by using its independent and strong 552.9 keV (100\%) gamma-line and the data from an activity measurement series with very short cooling times. The results are cumulative cross sections and are presented in Fig. 3 together with the results of the previous experimental works and the curves calculated by the nuclear reaction model codes TALYS 1.8 (TENDL-2015) and EMPIRE 3.2. Our results are in very good agreement with the previous data of Rebeles et al. \cite{19} in the overlapping energy region, but our work extends the energy range of the data up to 50 MeV. Both model codes follow the trend of the experimental data, but EMPIRE 3.2 strongly overestimates, while TENDL-2015 strongly underestimates above 32 MeV. The estimation of EMPIRE 3.2 is acceptable up to 32 MeV.

\begin{figure}
\includegraphics[scale=0.3]{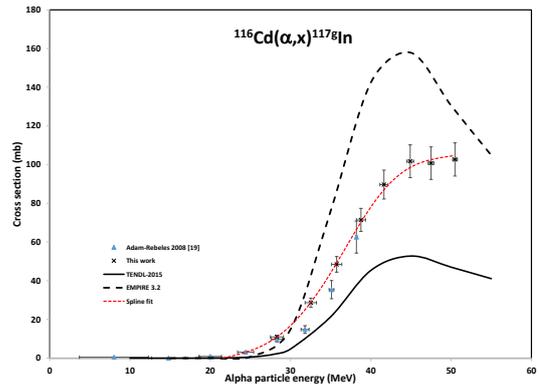}
\caption{Excitation function of the $^{116}$Cd($\alpha$,x)$^{117g}$In nuclear reaction in comparison with the previous literature data and the results of the theoretical nuclear reaction model codes EMPIRE 3.2 and TALYS 1.8 (TENDL-2015)}
\end{figure}

\subsubsection{$^{116}$Cd($\alpha$,x)$^{116m2}$In reaction}
\label{4.1.4}
The $^{116}$In has a very short-lived (T$_{1/2}$ = 2.18 s) meta-stable state, a longer-lived (T$_{1/2}$ = 54.29 min) meta-stable state and a very short-lived (T$_{1/2}$ = 14.1 s) ground state. All levels decay finally to the stable $^{116}$Sn. No mother isotope contributes by decay to the production of this radionuclide. So this isotope can be produced independently. In this measurement only the cross section for the longer-lived isomeric state could be assessed by using its relatively strong gamma-lines above 1 MeV. Having no interferences, no corrections were necessary and the data were taken from the measurement series with the best statistics. Our results, together with the previous experimental data and the results of the theoretical nuclear model code calculations are seen in Fig. 4. Our results are in acceptable agreement with the previous data of Adam-Rebeles et al. \cite{19} but extend the energy range of the available data. EMPIRE 3.2 and TENDL-2015 follow the experimental results with an acceptable agreement. According to Fig. 4 the estimation of EMPIRE 3.2 can be considered as better, especially in the energy region above 40 MeV.

\begin{figure}
\includegraphics[scale=0.3]{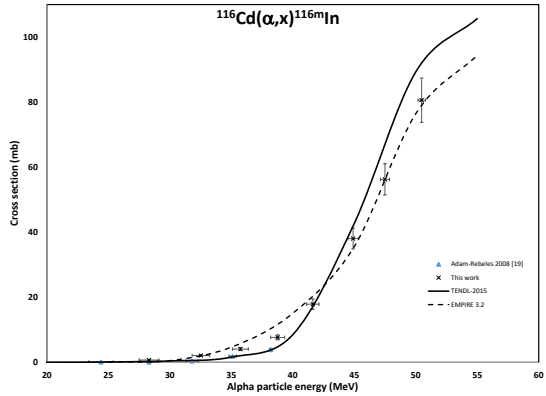}
\caption{Excitation function of the $^{116}$Cd($\alpha$,x)$^{116m}$In nuclear reaction in comparison with the previous literature data and the results of the theoretical nuclear reaction model codes EMPIRE 3.2 and TALYS 1.8 (TENDL-2015)}
\end{figure}

\subsubsection{$^{116}$Cd($\alpha$,x)$^{115m}$In reaction}
\label{4.1.5}
$^{115}$In has a very long-lived ground-state (T$_{1/2}$ = 4.41 $10^{14}$ a) and a meta-stable stable (T$_{1/2}$ = 4.486 h). It can be produced directly through the ($\alpha$,p4n) reaction and it is also fed by the decay of the much longer-lived $^{115m}$Cd mother isotope’s isomeric state (3\%). To separate the contribution of the mother we made corrections and we have taken the data from a measurement series with the best statistics, in such a way the decay contribution was kept under 3\%. Our results are presented in Fig. 5 together with the data from the nuclear reaction model code calculations. No previous experimental data were found in the literature. Our data are a bit scattered and have relatively large uncertainties because of the lower statistic (short measuring time after short cooling time). The trend of both EMPIRE 3.2 and TENDL-2015 corresponds to our experimental data, but the estimation of EMPIRE 3.2 is better. 

\begin{figure}
\includegraphics[scale=0.3]{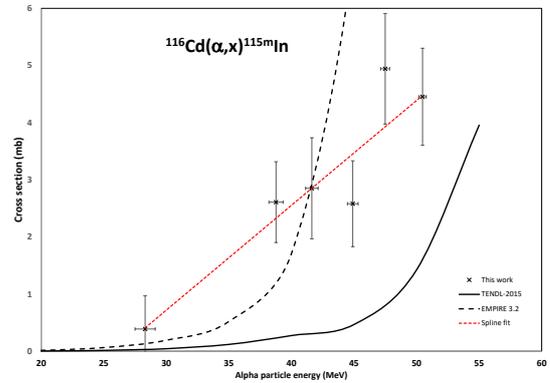}
\caption{Excitation function of the $^{116}$Cd($\alpha$,x)$^{115m}$In nuclear reaction in comparison with the results of the theoretical nuclear reaction model codes EMPIRE 3.2 and TALYS 1.8 (TENDL-2015)}
\end{figure}

\subsubsection{$^{116}$Cd($\alpha$,x)$^{115m}$Cd reaction}
\label{4.1.6}
$^{115}$Cd has a relatively long-lived ground state (T$_{1/2}$ = 53.46 h) and a much longer-lived isomeric state (T$_{1/2}$ = 44.56 d). We could assess cross section data for both states. The $^{115}$Cd can be produced directly through the ($\alpha$,2p3n) reaction including the lower threshold complex particle emission channel ($\alpha$,$\alpha$n) and d and t particles in the outgoing channel. The production of the $^{115m}$Cd can also be contributed by the $^{115}$Ag short-lived mother isotope, but in a very low percentage. To let the mother isotope decay completely our presented data were taken from a measurement series after relatively long cooling times. Our cumulative data are shown in Fig. 6. together with the results of previous experiments from the literature and the results of the theoretical nuclear model code calculations. Our data are in good agreement with the previous data of Montgomery and Porile \cite{20} and give energy range extension to it. Both TENDL-2015 and EMPIRE 3.2 produce a maximum on the calculated curves at slightly different energies, which could not be confirmed by any of the experimental results. Both model curves show very different trends. 

\begin{figure}
\includegraphics[scale=0.3]{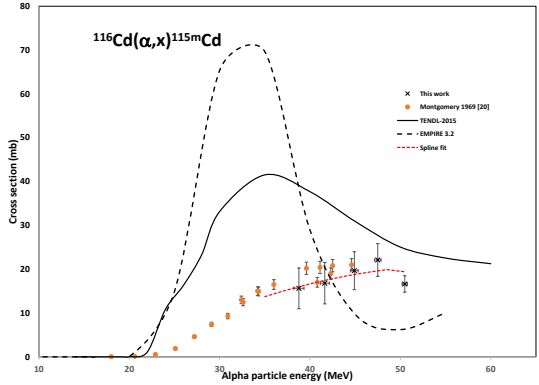}
\caption{Excitation function of the $^{116}$Cd($\alpha$,x)$^{115m}$Cd nuclear reaction in comparison with the previous literature data and the results of the theoretical nuclear reaction model codes EMPIRE 3.2 and TALYS 1.8 (TENDL-2015)}
\end{figure}

\subsubsection{$^{116}$Cd($\alpha$,x)$^{115g}$Cd reaction}
\label{4.1.7}
$^{115m}$Cd decays by 97\% to the ground state of $^{115g}$In and the rest to its exited level $^{115m}$In, so it has no contribution to the cross section of its ground state $^{115g}$Cd. It can only be fed from the short-lived $^{115}$Ag, so in this context our cross section is cumulative. The activities were determined by using the 527.9 keV (27.5\%) independent gamma-line from a series with longer cooling times in order to let the $^{115}$Ag levels decay. The results are shown in Fig. 7 together with the previous experimental results from the literature and with the results of the nuclear model code calculations. Our results are in acceptable agreement with the previous experimental data of Adam-Rebeles et al. \cite{19} and Montgomery and Porile \cite{20}. TENDL-2015 and EMPIRE 3.2 give very similar trend and also similar values, but the local maximum of the excitation function suggested by them cannot be confirmed by the experimental results.

\begin{figure}
\includegraphics[scale=0.3]{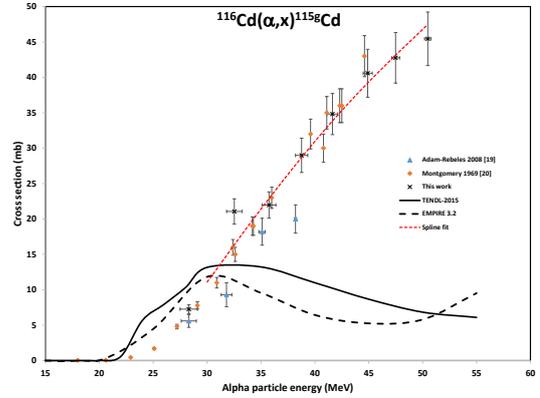}
\caption{Excitation function of the $^{116}$Cd($\alpha$,x)$^{115g}$Cd nuclear reaction in comparison with the previous literature data and the results of the theoretical nuclear reaction model codes EMPIRE 3.2 and TALYS 1.8 (TENDL-2015)}
\end{figure}

\begin{table*}[t]
\tiny
\caption{Experimental cross sections of the $^{117m}$Sn, $^{117m,g}$In, $^{116m}$In, $^{115m}$In and $^{115m,g}$Cd}
\centering
\begin{center}
\begin{tabular}{|p{0.2in}|p{0.2in}|p{0.2in}|p{0.2in}|p{0.2in}|p{0.2in}|p{0.2in}|p{0.2in}|p{0.2in}|p{0.2in}|p{0.2in}|p{0.2in}|p{0.2in}|p{0.2in}|p{0.2in}|p{0.2in}|} \hline 
\multicolumn{2}{|p{0.4in}|}{\textbf{Bombarding energy\newline E $\pm\Delta$E\newline MeV}} & \multicolumn{14}{|p{2.8in}|}{\textbf{Cross section $\sigma \pm \Delta\sigma$ \newline mb}} \\ \hline 
\multicolumn{2}{|p{0.4in}|}{} & \multicolumn{2}{|p{0.4in}|}{$^{117m}$Sn} & \multicolumn{2}{|p{0.4in}|}{$^{117m}$In} & \multicolumn{2}{|p{0.4in}|}{$^{117g}$In} & \multicolumn{2}{|p{0.4in}|}{$^{116m}$In} & \multicolumn{2}{|p{0.4in}|}{$^{115m}$In} & \multicolumn{2}{|p{0.4in}|}{$^{115g}$Cd} & \multicolumn{2}{|p{0.4in}|}{$^{115m}$Cd} \\ \hline 
50.5 & 0.3 & 178 & 15 & 6.7 & 1.3 & 103 & 9 & 81 & 6.8 & 4.5 & 0.8 & 45 & 4 & 17 & 2 \\ \hline 
47.5 & 0.4 & 250 & 21 & 5.5 & 1.2 & 101 & 8 & 56 & 4.8 & 4.9 & 1.0 & 43 & 4 & 22 & 4 \\ \hline 
44.9 & 0.4 & 366 & 30 & 6.5 & 1.0 & 102 & 9 & 38 & 3.2 & 2.6 & 0.8 & 41 & 3 & 20 & 4 \\ \hline 
41.6 & 0.5 & 582 & 48 & 5.3 & 0.9 & 90 & 8 & 18 & 1.6 & 2.9 & 0.9 & 35 & 3 & 17 & 5 \\ \hline 
38.8 & 0.6 & 843 & 70 & 5.2 & 0.9 & 72 & 6 & 8 & 0.8 & 2.6 & 0.7 & 29 & 2 & 16 & 5 \\ \hline 
35.7 & 0.6 & 1073 & 89 & 5.6 & 0.8 & 49 & 4 & 4.1 & 0.4 &  &  & 22 & 2 &  &  \\ \hline 
32.5 & 0.7 & 1178 & 98 & 3.7 & 0.6 & 29 & 2 & 2.1 & 0.2 &  &  & 21 & 2 &  &  \\ \hline 
28.3 & 0.8 & 852 & 71 & 2.8 & 0.5 & 11 & 1 & 0.7 & 0.1 & 0.4 & 0.6 & 7.3 & 0.6 &  &  \\ \hline 
\end{tabular}

\end{center}
\end{table*} 

\subsection{Thick target yields}
\label{4.2}
The thick target yields of the most important radioisotopes from this experiment ($^{117m}$Sn, $^{117m,g}$In, $^{116m}$In, $^{115m}$In and $^{115m,g}$Cd) were determined by using a spline fit to the experimental cross section data points and using  numerical integration over the required energy range taking into account the stopping of the alpha particles in the target material. The calculated physical yield curves are presented in Fig. 8 together with the available experimental data from the literature, measured on enriched $^{116}$Cd target. 
For $^{117m}$Sn production a previous result was reported by Maslov et al. \cite{32}. They measured on 95\% enriched $^{116}$Cd target with a thickness of 0.01107 g/cm2, by using 35 MeV bombarding energy. Under these conditions they measured a thin target yield of 410 kBq/$\mu$Ah, which corresponds to 0.1139 GBq/C. By using our yield data, we have also calculated thin target yield with the given thickness and bombarding energy of Maslov et al. The calculation resulted in a thin target yield of 0.1178 GBq/C, which is in a good agreement with the Maslov’s result. Both values are shown in Fig. 8. Except for $^{117m}$Sn, no experimental literature yield data were found on alpha particle induced reaction on enriched $^{116}$Cd. The good agreement with the Maslov’s results further confirms that our experiments, measurements, evaluation and interpretation were correct for the $^{117}$Sn isotope, and also confirms the experiments and measurements for the rest of the measured radioisotopes. Extrapolating the excitation function of $^{117m}$Sn the recommended energy range for production is 56 to 25 MeV. In this energy window the expected thick target yield is 1.75 GBq/C. A possible long-lived contaminant is the $^{115m}$Cd, its expected yield in the above energy range is 10.6 MBq/C and the activity ratio (0.6\% after 2 hours irradiation) increases with the time, because the contaminant has longer half-life than the $^{117m}$Sn isotope. Another possible long-lived contaminants are the $^{119m}$Sn (T$_{1/2}$ = 293.1 d) and $^{113}$Sn (T$_{1/2}$ = 115.09 d). The production of 113Sn from $^{116}$Cd is avoided because of the threshold energy (56.34 MeV) of the ($\alpha$,7n) reaction. On the 2-3 order of magnitude less $^{114}$Cd the ($\alpha$,5n) reaction is still possible (E$_{th}$ = 41 MeV), but the contribution could not be measured and can be considered as negligible. The cross section of the $^{119m}$Sn production on $^{116}$Cd at 25 MeV is an order of magnitude lower than the $^{117m}$Sn producing reaction \cite{19}, and further decreases with the increasing bombarding energy. Its low energy photons could not be measured in our experiments. All the other possible contaminants have shorter half-lives, so the contamination ratios improve with the increasing cooling time. Those contaminants having much higher activity at the end of the irradiation (EOB) ($^{117m,g}$In, $^{116m}$In) practically disappear in a couple of days because of the much shorter half-life. The same is valid for the remaining short-lived and low yield radioisotopes. The produced $^{117m}$Sn is no-carrier added, its specific activity can be determined experimentally. 

\begin{figure}
\includegraphics[scale=0.3]{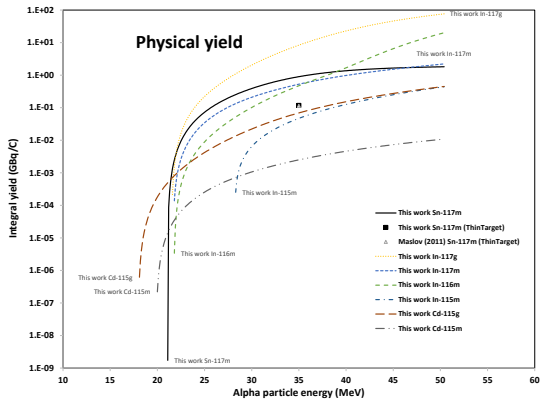}
\caption{Thick target physical yields of the investigated radioisotopes. Thin target yield of Maslov et al. is also presented in this figure and compared with the thin target yield calculated from our results using the same input parameters (target thickness, energy window, enrichment)}
\end{figure}

\section{Summary and conclusion}
\label{7}
This work presents a new series of measurements in order to determine the cross section and yield of the medically important theranostic radioisotope $^{117m}$Sn by alpha particle irradiation on enriched $^{116}$Cd target up to 51 MeV. As auxiliary results, the cross sections of several other radioisotopes co-produced on $^{116}$Cd or on minor stable Cd isotopes present in the target, were measured. The cross section of $^{115m}$In was measured for the first time. For all other radioisotopes we could give new cross section values between 40 and 51 MeV. In most cases a good agreement was found between our results and the previous data. The comparison with the previous experimental data from the literature in the overlapping energy region proved that our methods and calculations are correct and in such a way our values have also been confirmed. We have also compared our experimental data with the results of theoretical nuclear reaction model code calculations made by the TALYS 1.8 (TENDL-2015) and EMPIRE 3.2 (Malta)by using default input. The agreement between the model calculations and the experimental results was good only in some cases (e.g. $^{116m}$In in the whole investigated energy region), but the model codes could predict the trend of the excitation functions in most cases (e.g. $^{117g}$In). It must also be mentioned that in several cases, because of possible wrong calculation of complex particle emission the predictions of the codes completely failed (e.g. $^{117m}$In (EMPIRE), $^{115m}$Cd (both), $^{115g}$Cd ($>$ 35 MeV both)).

\section{Acknowledgements}

This work was performed at the RI Beam Factory operated by the RIKEN Nishina Center and CNS, University of Tokyo. This work was carried out in the frame of the standing HAS-JSPS (Hungary–Japan) bilateral exchange agreement. The authors acknowledge the support of the respective institutions in providing technical support and use of experimental facilities. (Contract No.: NKM-89/2014).
 



\bibliographystyle{elsarticle-num}
\bibliography{Cd116}







\end{document}